\documentclass[preprint,12pt]{elsarticle}

\usepackage{graphicx}
\usepackage{subfigure}

\usepackage{amssymb}

\def\MeV{\ifmmode {\mathrm{\ Me\kern -0.1em V}}\else
                   \textrm{Me\kern -0.1em V}\fi}
\def\GeV{\ifmmode {\mathrm{\ Ge\kern -0.1em V}}\else
                   \textrm{Ge\kern -0.1em V}\fi}
\def\cm{\ifmmode {\mathrm{\ c\kern -0.05em m}}\else
                   \textrm{c\kern -0.05em m}\fi}
\def\ns{\ifmmode {\mathrm{\ n\kern -0.05em s}}\else
                   \textrm{n\kern -0.05em s}\fi}
\def\mm{\ifmmode {\mathrm{\ m\kern -0.1em m}}\else
                   \textrm{m\kern -0.1em m}\fi}
\def\mV{\ifmmode {\mathrm{\ m\kern -0.1em V}}\else
                   \textrm{m\kern -0.1em V}\fi}
\journal{Nuclear Instruments \& Methods in Physics Research}

\begin{document}

\begin{frontmatter}



\title{Fast Beam Conditions Monitor BCM1F for the CMS Experiment
}


\author[cern,unigenf]{A. Bell}

\author[desy]{E. Castro}

\author[cern,wisc]{R. Hall-Wilton}

\author[desy]{W. Lange} 

\author[desy,btu]{W. Lohmann}

\author[cern]{A.~Macpherson} 

\author[desy,btu]{M. Ohlerich}

\author[zeeland]{N. Rodriguez} 

\author[cern]{V. Ryjov}

\author[desy,btu]{R.S. Schmidt} 

\author[rutgers]{R.L. Stone}
\address[btu]{Brandenburgische Technische Universit\"at, 03046 Cottbus, Germany}
\address[cern]{CERN, 1211 Geneva 23, Switzerland}
\address[desy]{DESY, 15738 Zeuthen, Germany}
\address[rutgers]{Rutgers University, 08854 Piscataway, NJ, USA}
\address[unigenf]{Universit\'e    de Gen\`eve, 1211 Geneva, Switzerland} 
\address[zeeland]{Canterbury University, 8041 Christchurch, New Zealand}
\address[wisc]{University of Wisconsin, Madison, WI 53706-1481, USA}
\begin{abstract}
The CMS Beam Conditions and Radiation Monitoring System, BRM,
will support beam tuning, protect 
the CMS detector from 
adverse beam conditions, and
measure the accumulated dose close to or inside all sub-detectors.
It is composed of different sub-systems measuring either the particle flux near the
beam pipe with time resolution between nano- and microseconds or the integrated dose 
over longer time intervals.
This paper presents the Fast Beam Conditions Monitor,
BCM1F, which is designed for fast flux monitoring 
measuring both beam halo and 
collision products. BCM1F 
is located inside the CMS pixel detector volume close to the 
beam-pipe. It
uses sCVD 
diamond sensors and radiation hard front-end electronics, 
along with an analog optical readout of the signals.
The commissioning of the system and its successful operation during the first beams of the LHC 
are described.
\end{abstract}


\end{frontmatter}


\section{Introduction}
\label{intro}
The CMS experiment~\cite{CMS} at LHC~\cite{LHC} 
will be situated in an unprecedentedly high 
radiation field.
The LHC is designed to run with 362 MJ 
of stored energy in one beam and with proton intensities 
of more than 10$^{14}$ per beam. 
These beams will generate a continuous flux of halo particles 
near the beam-pipe and when colliding also interaction secondaries, predominantly at small polar angles.
The collected dose will be largest for the innermost detectors, which are therefore    
designed  
with very high radiation 
tolerance. 

However, also short term
losses 
of the beams may cause serious 
damage to detector elements, in particular
to front-end electronics due to large ionisation.
In addition,
the innermost detectors need a sufficiently low occupancy level 
for successful data taking.   

To monitor the particle fluxes near the beam-pipe and the radiation level in the sub-detectors,
beam conditions and radiation monitors, BRM~\cite{CMS,BCM1}, are installed in the CMS detector. 
Several slow systems of BRM will be used to measure the
accumulated dose near the volume of all sub-detectors. These measurements
are necessary to understand potential 
longer term damage to detector elements.
  
Particle flux monitors are installed close to the beam-pipe. 
Two such monitors measure the integral particle flux over
half an orbit or over a bunch train using the signal current
in polycrystalline small 
diamond sensors and integrate it
over 
about 40 $\mu$s and 5 $\mu$s,
respectively. 
These fluence measurements
will assist beam tuning, indicate critical beam halo conditions
for the inner detectors and
initiate LHC beam aborts when
conditions are such that detectors might be endangered.
These systems are described in detail in Ref.~\cite{abell}.
Other LHC experiments also installed beam condition monitors using polycrystalline
diamond sensors~\cite{other_experiments}.

The Fast Beam Conditions Monitor, hereafter
referred to as BCM1F, will be sensitive to very fast changes of the beam conditions
and provide diagnostics with a time resolution better than the time between
bunch crossings, hence, for example, being able to 
flag problematic beam conditions resulting in bursts
of beam loss over very short periods of time.  
Such beam losses are considered to be one of the principle 
damage scenarios for CMS detector components.
In addition, it will store real-time data 
to allow post-mortem analyses in the case of 
beam accidents.

\section{BCM1F System Overview}
\label{over}

\begin{figure}[htb]
\centering
\includegraphics[width=4in]{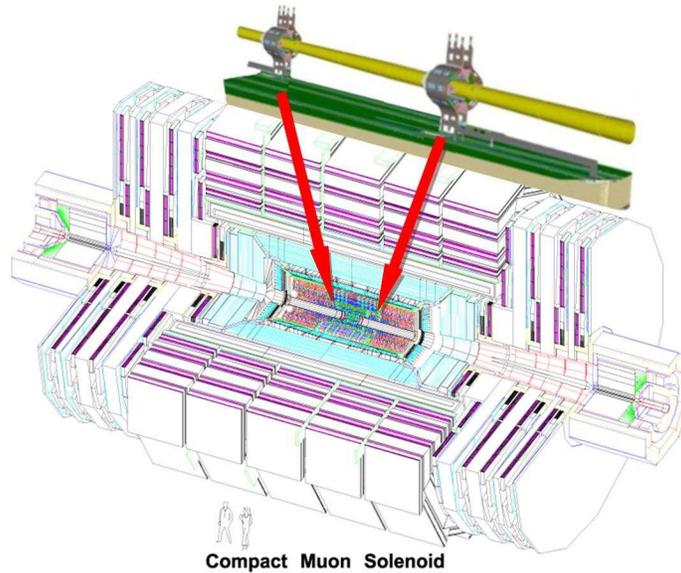}
\caption{A sketch of the CMS detector. The positions of the BCM1F in front of the
pixel detector and inside the tracker are indicated by the arrows.
At both sides of the IP a carbon fibre structure around the beam-pipe, as shown in the 
upper part, supports the sensor modules.}
\label{fig:CMS_detector}
\end{figure}

BCM1F uses single-crystal 
CVD\footnote{Chemical Vapor Deposition} 
diamond sensors, hereafter denoted as sCVD, for particle detection. 
Sensors made of
sCVD are sufficiently fast to match the time resolution requirements, 
and small 
enough to be inserted into areas close to key detector components 
without adding substantial material or services.  
Four sCVD sensors, each with a volume of 
5$\times$5$\times$0.5$\mm^3$, are positioned in a plane perpendicular to the beam-pipe 
on each side of the 
IP at a distance of about 1.8 m and   
at a radius of 4.5$\cm$ from 
the nominal beam position, as sketched in Figure~\ref{fig:CMS_detector}.
The position of the BCM1F is chosen to 
be optimal in  terms of time separation between 
ingoing and outgoing particles from the IP. Relativistic particles
need about 6$\ns$ to move between one of the detector planes and the IP. 
Hence, gated rate measurements of the BCM1F will allow to separate the fluxes
from both beam halo of each direction and interactions products.

\begin{figure}[htb]
\centering
\includegraphics[width=5.2in]{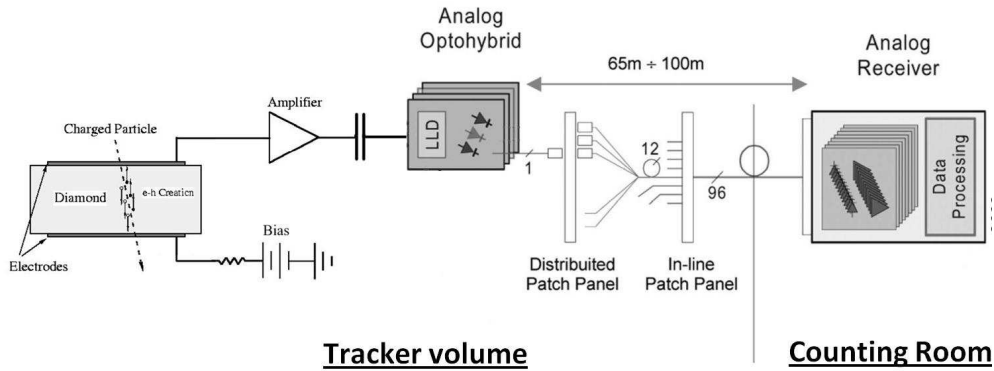}
\caption{
The readout scheme of BCM1F. The signal generated in the sensor
is amplified and shaped in a JK16 preamplifier. The analog
opto-hybrid drives a 
laser for analog signal transmission via a single-mode fibre. The signals are digitized 
and processed in the counting room. 
}
\label{fig:readout_scheme}
\end{figure}

Each sensor is connected to a radiation hard preamplifier.
Its output signal is transmitted to the counting room 
over an analog optical link as shown schematically
in Figure~\ref{fig:readout_scheme}. 
Since neither cooling nor slow 
control equipment are available at the mounting positions,
the modules must be operated with low power dissipation and 
should work over long periods without a 
re-adjustment of the calibration parameters.

At the back-end of the readout in the 
counting room signals are digitized 
and processed in a PC. 
Flash ADCs, scalers and
multi-hit recording
TDCs
allow e.g. to monitor counts as a function of time
over an orbit.
Test-pulses are used to check the functionality of the system. 
Rates, multiplicities, timing and 
coincidence information are monitored and stored  
independently of the CMS data 
acquisition.

\section{sCVD Sensors}
\label{sCVD}
Outstanding properties, such as 
 a very low leakage current with negligible temperature 
dependence, a fast signal response and  
radiation hardness, make CVD diamond sensors 
attractive for the locations close to the interaction 
region. 
In previous experiments
polycrystalline 
diamond sensors have been successfully used as beam
conditions monitors~\cite{CDF,Babar} by measuring the currents 
created in the sensor by the crossing particles. 
However, integration over 
a certain time limits the time resolution of such devices.
In addition, due to crystal defects the
charge collection efficiency of polycrystalline CVD sensors is below 50\%
which may result in a signal-to-noise ratio not sufficient for the 
detection of minimum ionising particles, MIPs.

Here single crystal CVD diamond sensors
are used. They are characterised by nearly 100\% charge
collection efficiency and allow to count MIPs. 
They are operated as solid state 
ionisation chambers by applying 
high voltage to thin metal plates  
on both sides of a sensor to create an electric field in the bulk, as shown in 
Figure~\ref{fig:readout_scheme}.
Signals from crossing charged particles
are created due to the drift of electrons and holes released in the
bulk material.

The sCVD sensors are 
of 5$\times$5$\mm^2$ area and 500 $\mu$m thickness. 
They have been 
manufactured by Element Six~\cite{E6} after a few 
years of development and research in 
collaboration with the CERN RD42 project~\cite{RD42}.
A first application of an sCVD diamond sensor in a collider experiment 
was described in Ref.~\cite{zeus}.
\begin{figure}[htb]
\centering
\subfigure{
\includegraphics[width=6.5cm,height=5cm]{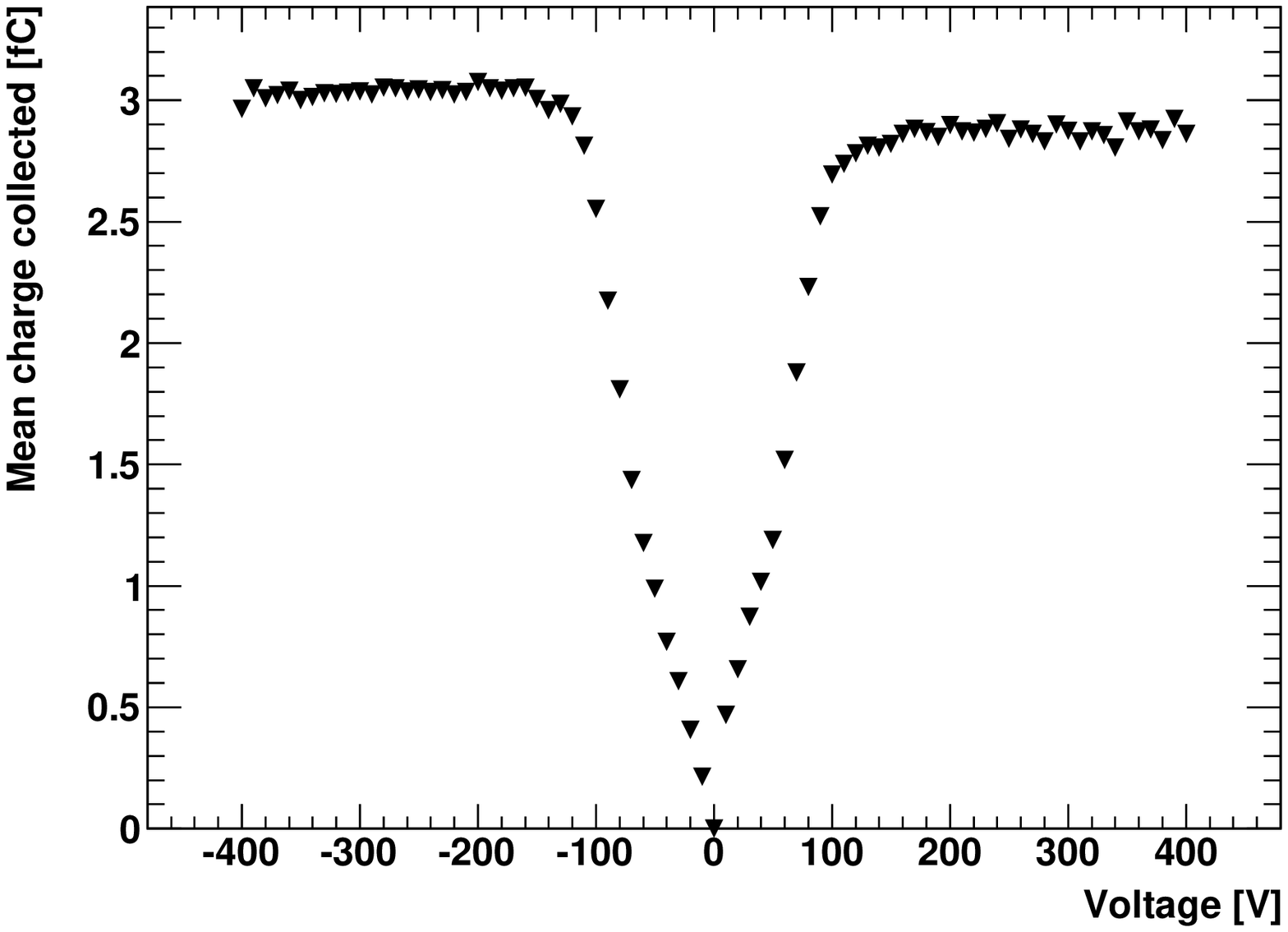}
}
\subfigure{
\includegraphics[width=6.5cm,height=5cm]{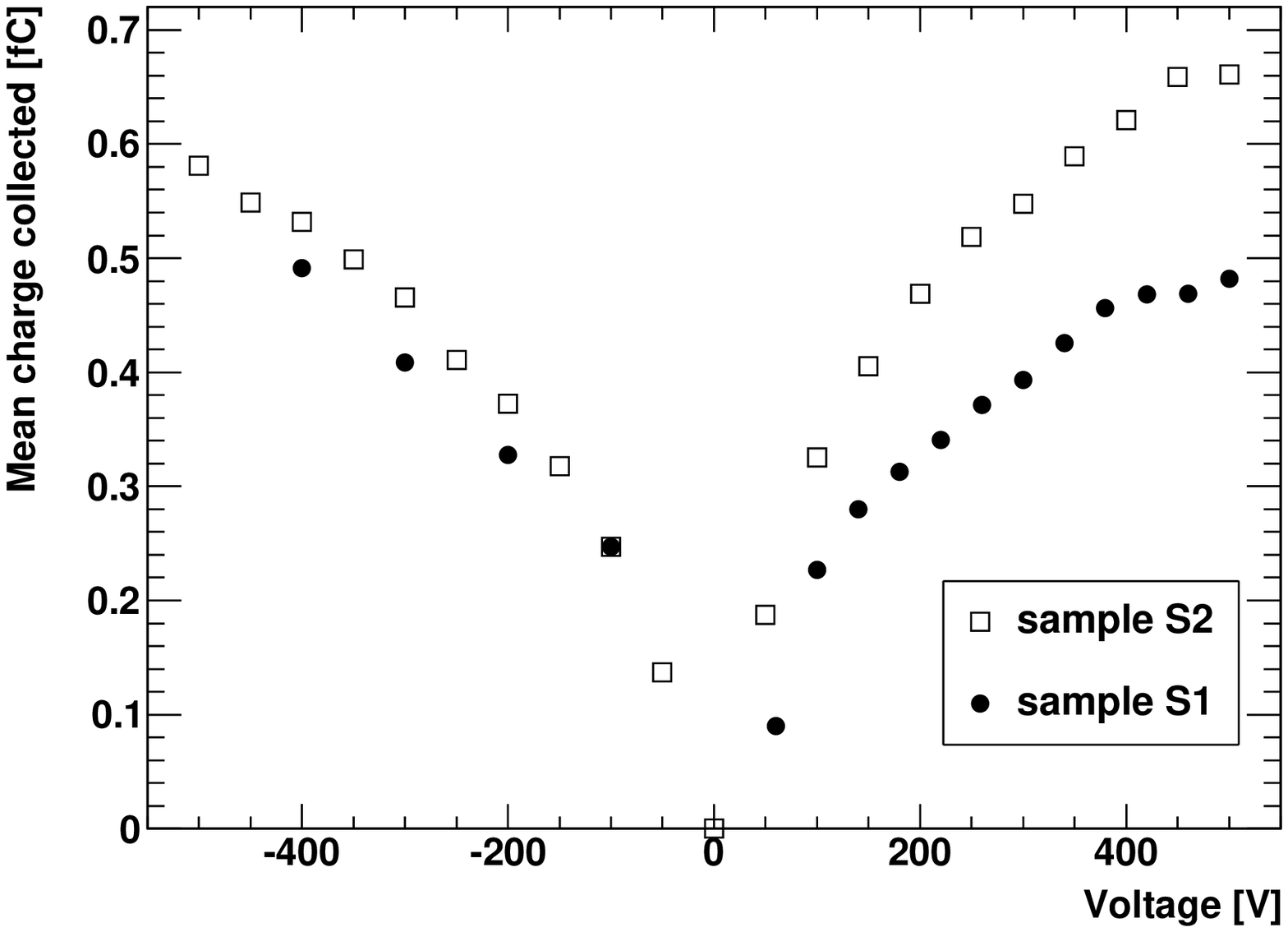}
}
\caption{
The signal amplitude as a function of the bias voltage
for a non-irradiated sCVD sensor (left) and for the two sensors after irradiation
(right).}
\label{fig:cvd_characteristics}
\end{figure}
\begin{figure}[htb]
\centering
\includegraphics[width=4in,height=6cm]{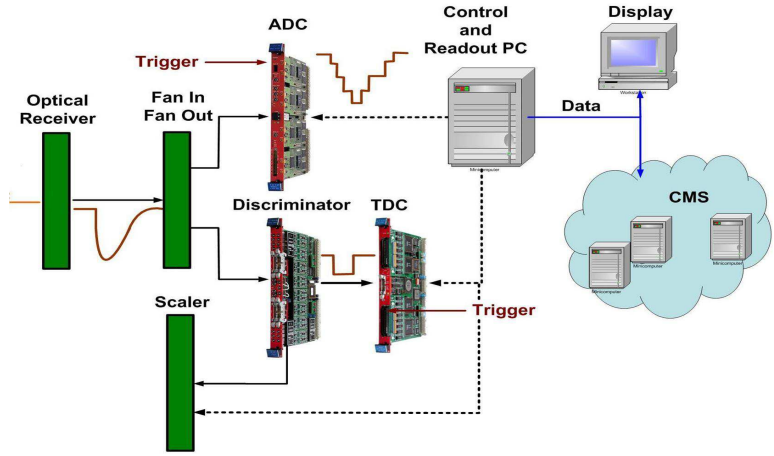}
\caption{
The schematic of the back-end readout. 
}
\label{fig:backend}
\end{figure}
\begin{figure}[htb]
\centering
\includegraphics[width=4in]{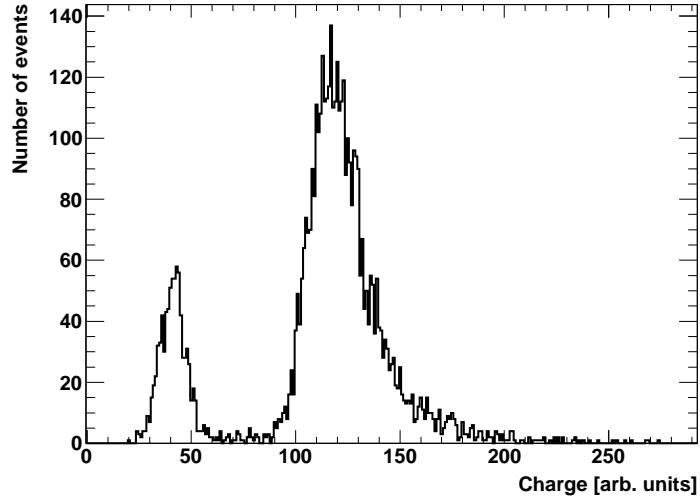}
\caption{
The spectrum of signals from relativistic electrons of a 
$^{90}$Sr source taken with a fully assembled BCM1F module read
out with a charge-integrating ADC. 
}
\label{fig:flash_spectrum}
\end{figure}   
\begin{figure}[htb]
\centering
\includegraphics[width=4in]{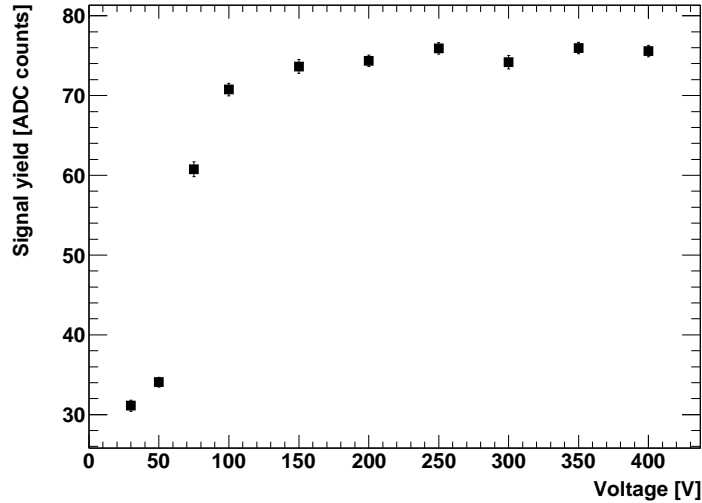}
\caption{
The peak value of the signal spectrum as a function of the bias voltage. 
}
\label{fig:signal_vs_voltage}
\end{figure}
\begin{figure}[htb]
\centering
\includegraphics[width=4in]{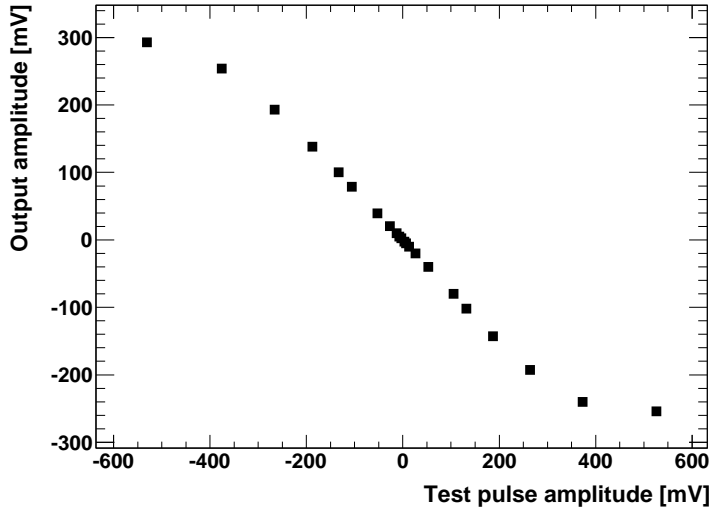}
\caption{
The size of the analog pulse measured at the opto-receiver module output
as a function of the amplitude of the test-pulse fed into the preamplifier.
}
\label{fig:linearity}
\end{figure}
\begin{figure}[htb]
\centering
\includegraphics[width=3.5in]{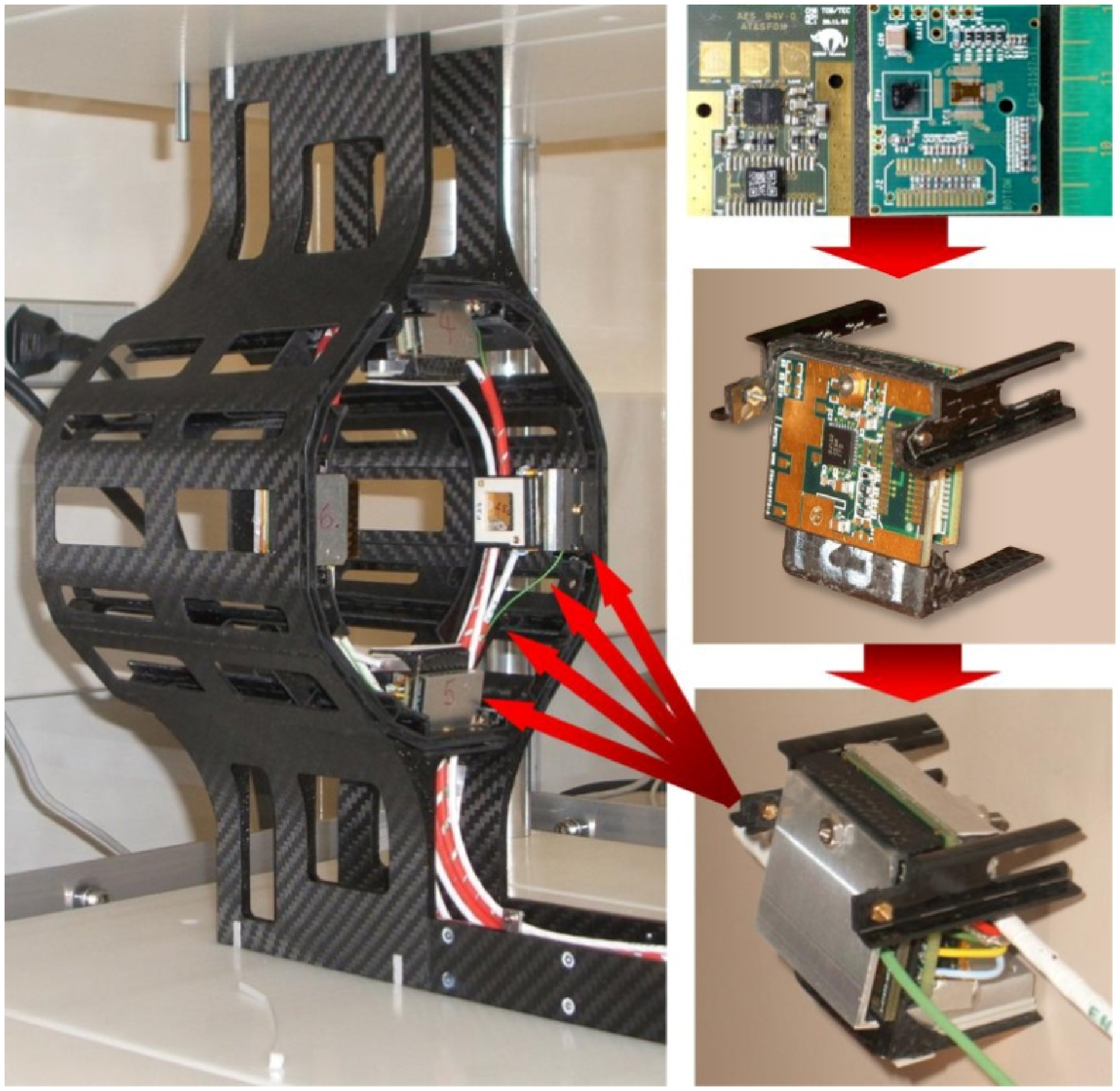}
\caption{
Left side: The carbon fibre carrier structure of BCM1 with the four modules installed.
Right side: Each module contains the sCVD sensor carrier (top left) and the preamplifier with the laser driver
(top right), which are combined to a sandwich (middle) and then connected with cables and optical fibres
and protected by an aluminum cage (bottom).
}
\label{fig:full_module}
\end{figure}

\section{Sensor Tests}
\label{test}

Before installation the sensors were tested in the laboratory.
The leakage currents and signal response to electrons from a $^{90}$Sr 
source were measured for all diamond sensors before assembly.
The leakage currents of the sensors are in the range of a few pico-Amperes. 
The signal amplitude, expressed as the average
collected charge, is shown as a function 
of the bias voltage for both polarities in   
Figure~\ref{fig:cvd_characteristics} for one of the sensors. 
The signal amplitude increases for increasing bias voltage up to about 120 V, 
corresponding to a field strength of about 0.25 V$\mu$m$^{-1}$,
and is constant thereafter.
The measurements for the other sensors show a very similar 
behavior. In a few cases slight differences in the signal size 
for different bias voltage polarities are observed. 
These sensors  
are operated with the polarity of the bias voltage 
giving the maximum signal yield.         

Two sample sensors were irradiated in a 60$\MeV$ proton beam at PSI up to a 
fluence of about 3$\times$10$^{14}$ 
protons per cm$^2$, corresponding to a 
fluence of 17.5$\times$10$^{14}$ MIPs per $\cm^2$~\cite{rad-hard_studies}.
The signal amplitude obtained from electrons 
of a $^{90}$Sr source is shown in 
Figure~\ref{fig:cvd_characteristics} for two sensors after irradiation.
It drops to about 20\% of the 
one measured with a non-irradiated sensor.
Whereas in the non-irradiated sample the signal amplitude saturates already at about
0.25 V$\mu$m$^{-1}$, in the irradiated sample no saturation is seen up to 
1~V$\mu$m$^{-1}$.
Comparing these results to previous studies~\cite{rad-hard_studies} 
of the performance of diamond sensors as a 
function of the fluence of 26$\MeV$ and 24$\GeV$ 
protons the hypothesis of enhanced 
damage at lower particle energies is supported.
The fluence investigated here is approximately that expected at the location of the BCM1F 
detector over the baseline LHC program.

\section{Readout Electronics}
\label{FE}
Each sensor is connected to a JK16 radiation hard amplifier 
ASIC~\cite{FE1}. 
The chip is 
fabricated in a commercial 0.25 $\mu$m CMOS technology 
hardened by appropriate layout techniques.

Each channel comprises a 
fast trans-impedance preamplifier with an active 
feedback loop 
and an amplifier-integrator 
stage with 20$\ns$ peaking time. An excellent 
noise performance is achieved by a careful adjustment 
of the feedback current through the gate voltage of the
feedback FET.
For a detector capacitance of 5 pF the measured noise 
amounts to about 700 electrons ENC in agreement with the specifications given in 
Ref.~\cite{FE1}. 
The measured charge gain is 20$\mV$/fC.

The analog signals are transmitted to the counting room
using an analog 
optical chain~\cite{optics1}
developed for the CMS tracker. The 
preamplifier's single-ended output is AC coupled to 
the custom-designed laser driver ASIC, which 
modulates the current of the edge-emitting laser diode. 
Single mode fibres from the pigtailed lasers are connected 
at the periphery of the tracker volume to an optical fan-in, which 
merges single fibres into a 12-fiber ribbon cable. In the counting 
room a corresponding ribbon connects directly to a 12-channel 
analog optical receiver card in a VME crate.  
 
\begin{figure}[htb]
\centering
\includegraphics[width=4in]{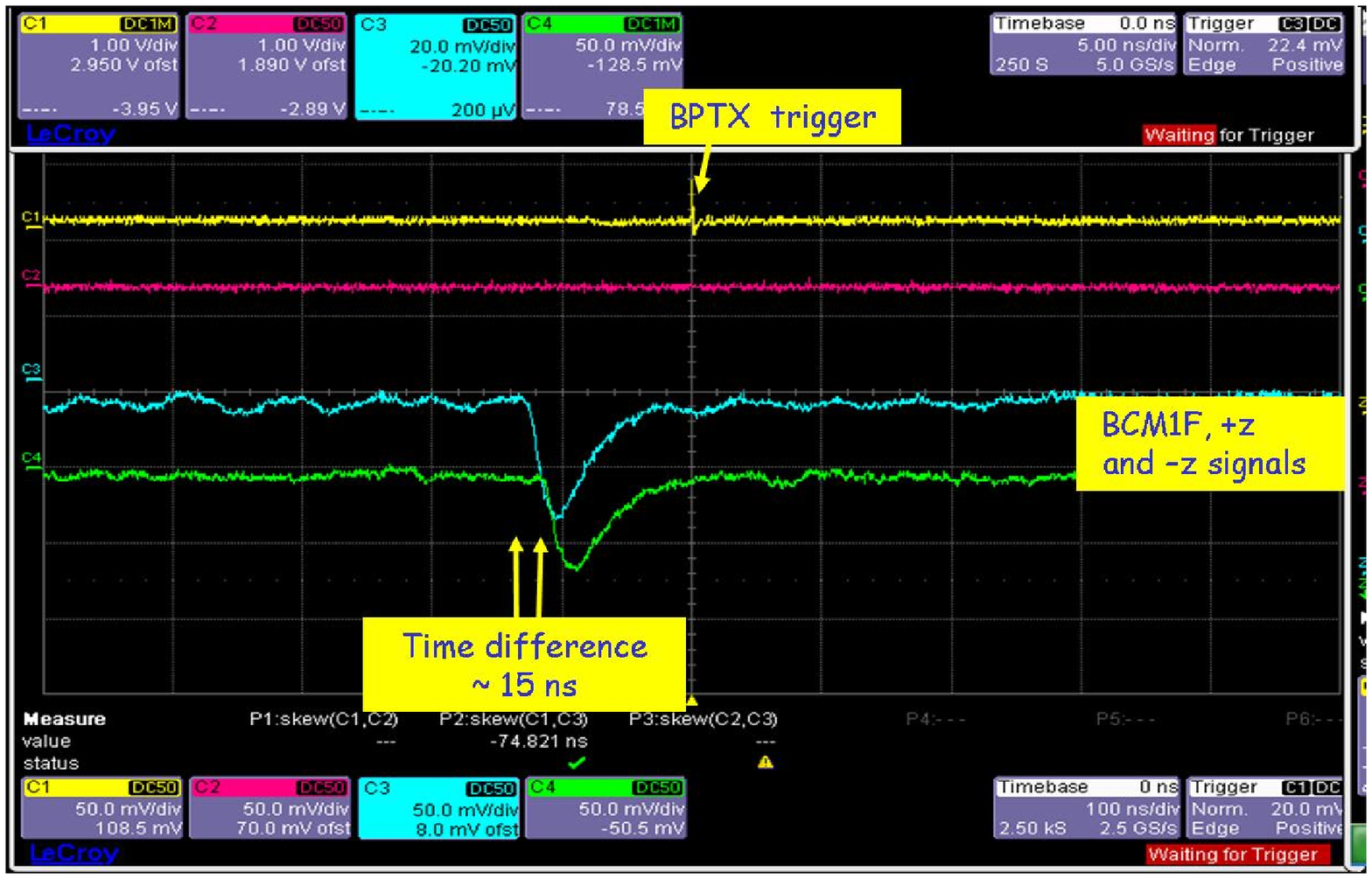}
\caption{
A snapshot of first signals in BCM1F from September 2008. The trigger was taken from the
BPTX bunch pickup (yellow line). The blue and the green signals 
result from an analog sum of the BCM1F signals of each side.
The shift of time between the two signals corresponds roughly to the time of flight of a relativistic particle
between both sensor planes.
}
\label{fig:pulse_scope}
\end{figure} 
\begin{figure}[htb]
\centering
\subfigure{
\includegraphics[width=6.5cm,height=5.5cm]{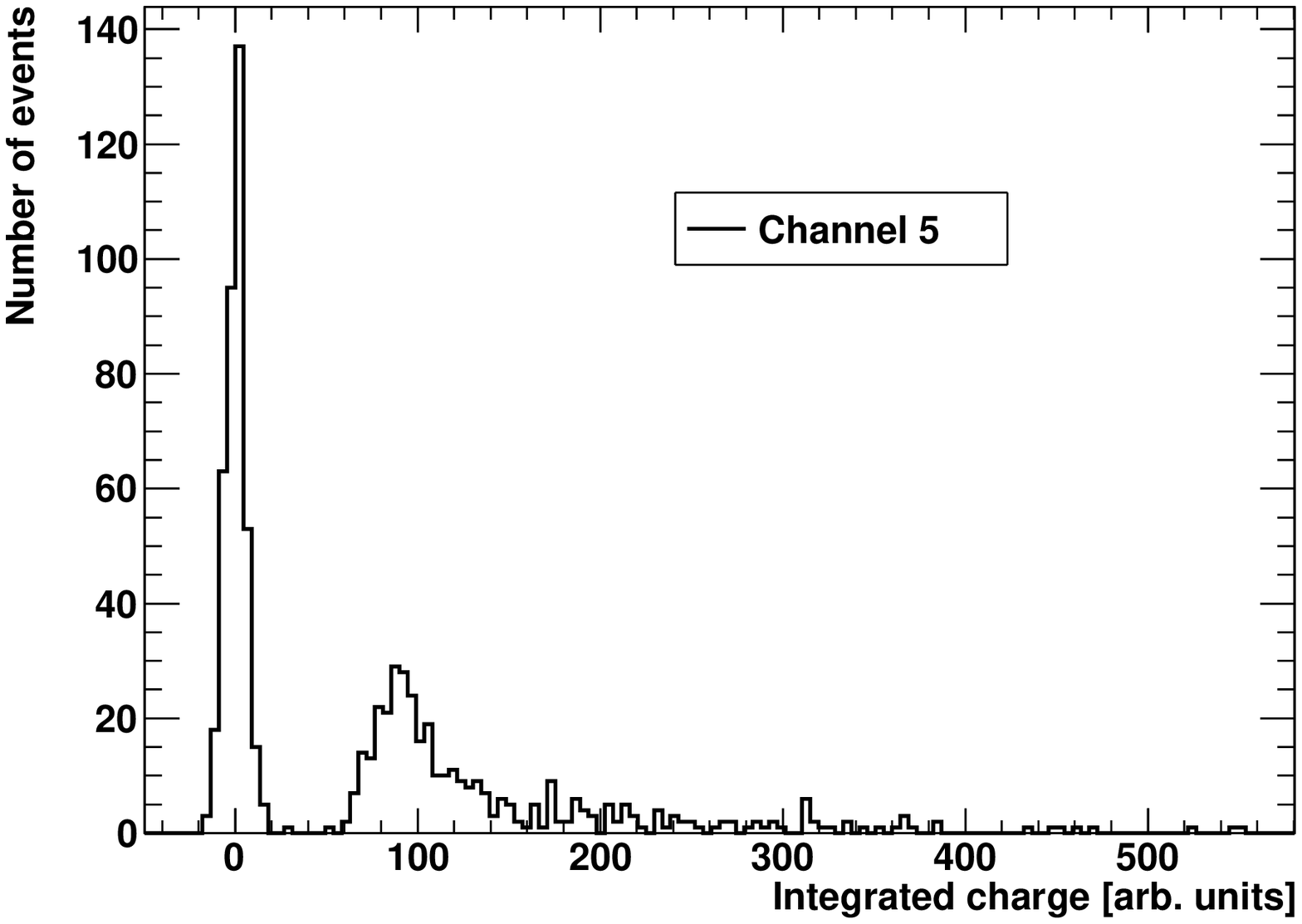}
}
\subfigure{
\includegraphics[width=6.5cm,height=5.5cm]{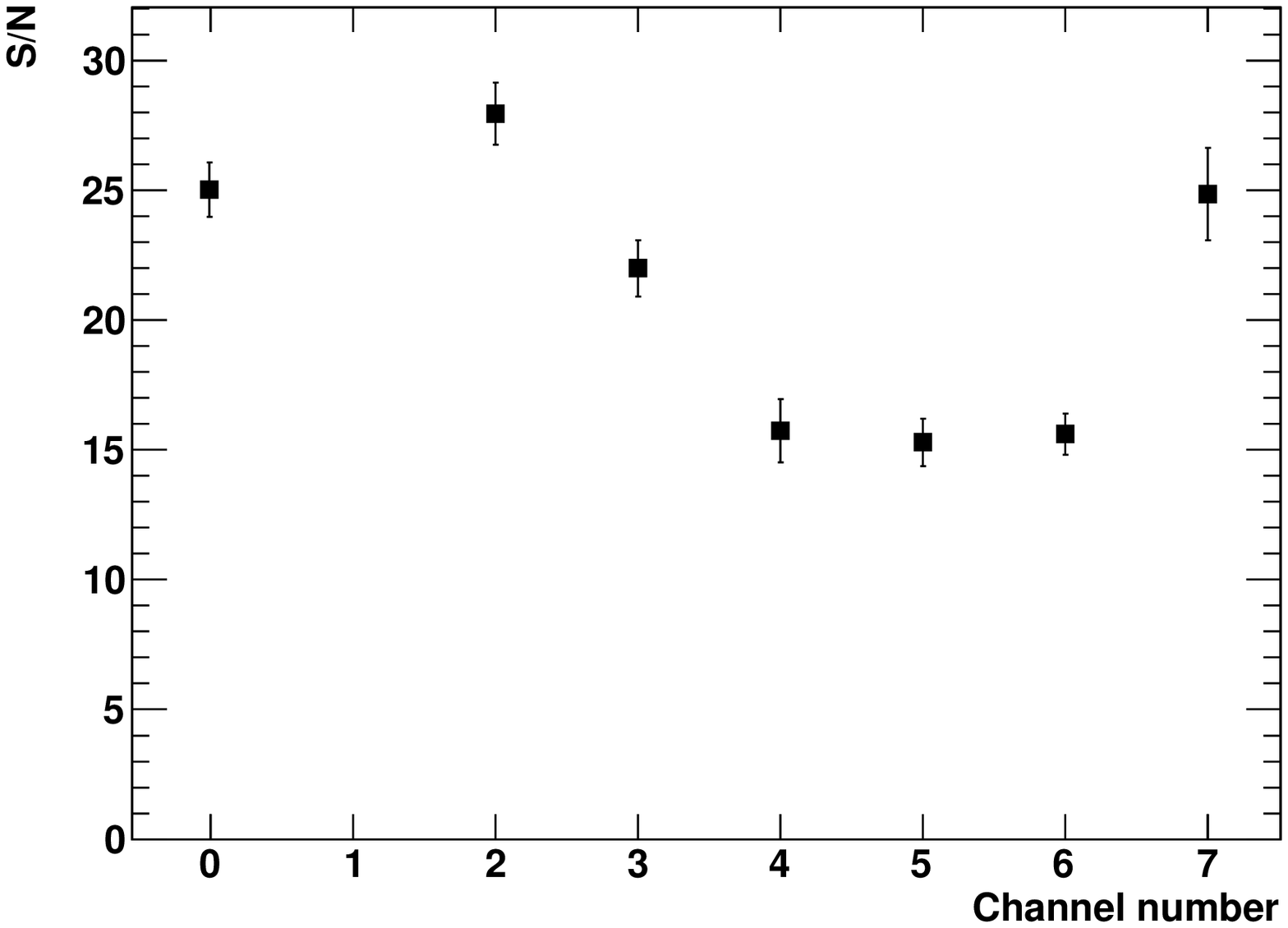}
}
\caption{
The distribution of the signals for particles crossing the sensor in one module of 
BCM1F (left)
and the signal-to-noise ratios
obtained from such distributions for each channel of BCM1F (right).
}
\label{fig:pulse_height}
\end{figure}
\begin{figure}[htb]
\centering
\includegraphics[width=3.5in]{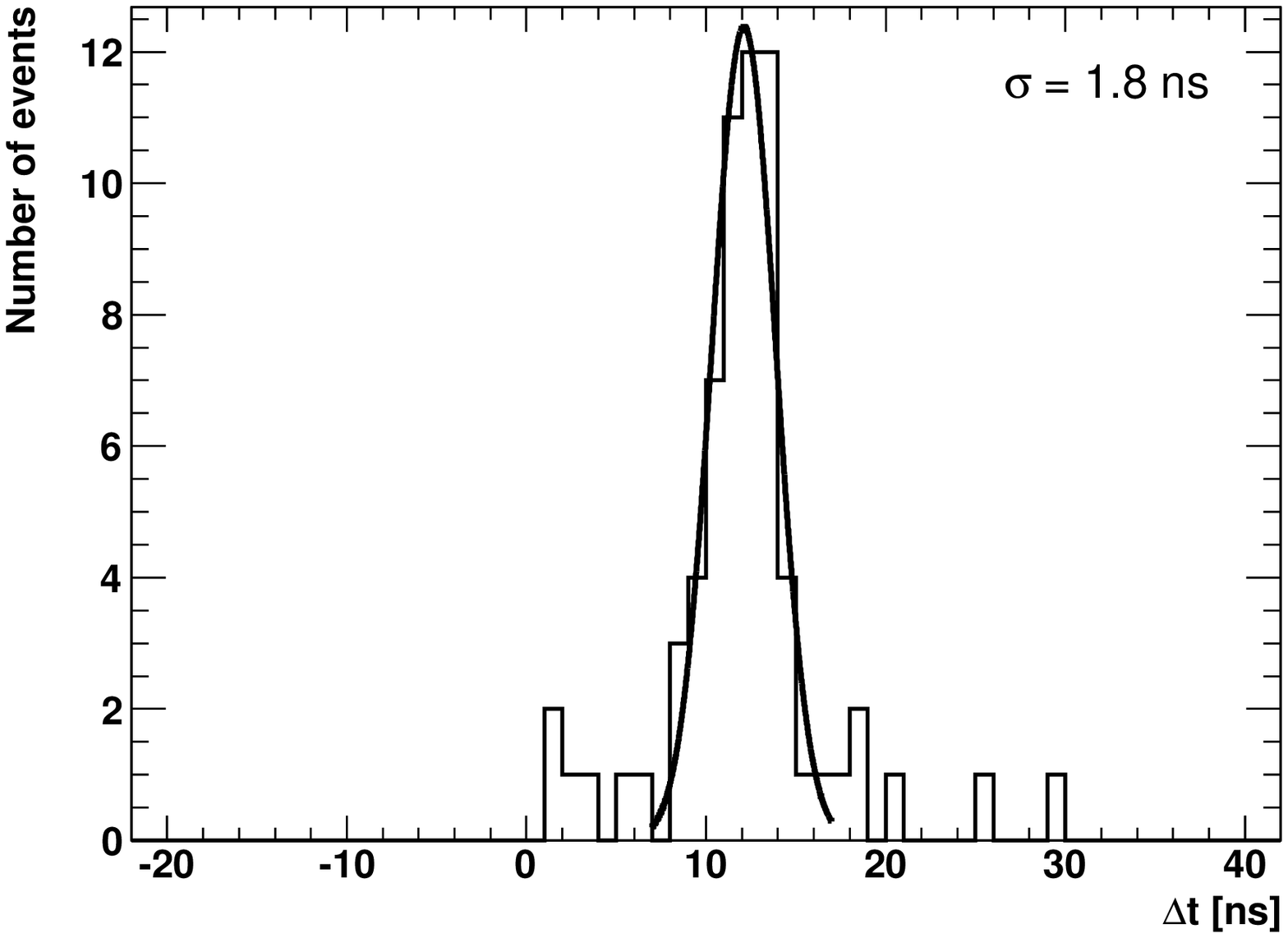}
\caption{
Distribution of the difference between arrival times of signals from sensors in
different z-planes but
 equal 
azimuthal angles. 
}
\label{fig:arrivel_time}
\end{figure}   
Minor modifications on the laser driver ASIC
board were done to allow mounting in two opposite 
orientations of the laser diode, required by the minimal 
bending radius of the pigtail fibres.  In contrast to the tracker 
application, for the BCM1F the gain and the 
laser diode bias current cannot be programmed via the 
foreseen I$^2$C interface.  Hence, attention 
was paid to choose the input polarity and the laser 
bias setting to preserve the dynamic range of the 
receiver side. In addition, the impacts of heat dissipation 
and the expected radiation dose on the  
laser diode performance degradation were taken into account. 
To ensure a small package size a piggy-back architecture  
was used for interconnecting and mounting the 
sensor, the amplifier and the analog optical hybrid
boards 
on their carriage.

At the back-end side of the readout 
the optical signals are converted into electrical signals
using an analog opto-receiver module.
Its output signals are distributed by analog fan-outs to ADC
inputs and to  
discriminators. 
A flash ADC performing 500 MS/s with 8 
bit resolution, V1721 from CAEN, is used to digitise the
signals.
This module can be triggered internally or externally. 
It can read out in full data mode up to 45 consecutive beam orbits or a 
corresponding number of user definable time intervals. 
Data is written into a ring buffer and tagged with time stamps.
It is read out via an on-board optical
link and processed in a PC. 

The discriminated signals are counted 
 in all channels 
with a V260 scaler from CAEN and used for on-line displays of hit rates. 
In addition, they are
digitised with
multi-hit capable 
TDCs V767 from CAEN
with 20 bit 
dynamic range and 0.8 ns-LSB resolution. The TDCs and the scalers are 
read out via a VME-bridge.
They will allow 
orbit-by-orbit counts to be obtained as a function of time for a detailed 
monitoring of beam halo and interaction products.  

Test-pulses are used to check the functionality of the system during operation.
The amplitude of the test-pulse induces a signal similar to one MIP in the preamplifier. 
A schematic of the complete back-end is shown in Figure~\ref{fig:backend}.

\section{Performance of the System Before Installation}
\label{performance}

The assembled front-end modules were tested before installation
using a  
$^{90}$Sr source. Relativistic electrons crossing the sensor
trigger a scintillation counter.
An example of a spectrum recorded with a charge-integrating ADC is shown
in Figure~\ref{fig:flash_spectrum}.
The distribution of the signal charge shows the expected Landau-shape.
The signal is clearly separated from the pedestal peak.
A signal-to-noise ratio of about 12 is estimated.
The values for the other channels are very similar.

These spectra were acquired for a range of increasing voltage applied across the sensor. 
Figure~\ref{fig:signal_vs_voltage} shows 
the most probable values 
of the pedestal-subtracted pulse 
height distributions measured as a function of the bias 
voltage. The 
maximum signal was reached, as expected, at an 
electric field of about 0.25 V$\mu$m$^{-1}$, corresponding to a bias voltage 
of 125 V for 
a sensor with 500 $\mu$m thickness. 
Above this voltage the signal is constant.

The linearity of the response of the whole readout chain
was investigated using test-pulses fed in a dedicated preamplifier
input. The result is shown 
in 
Figure~\ref{fig:linearity}.
For both polarities a linear response is found up to test-pulse 
amplitudes corresponding to approximately 
5 MIP equivalents. For test-pulse amplitudes above this value the readout 
becomes non-linear and approaches saturation at about 10 MIPs.

To test the proper functionality in 
the cooled tracker environment of CMS, 
the modules were operated in a climate chamber 
with five temperature cycles from $-$20 to $+$50 $^\circ$C.   
Leakage and supply currents as well as the test-pulse 
response were measured and found to be in the expected range. 
The stored results of these measurements will be compared
with measurements of the same quantities taken during operation of the 
modules in CMS.

\section{Installation in CMS}
\label{installation}

The components of the modules and the completed modules
mounted on the carrier structure are
shown in 
Figure~\ref{fig:full_module}.
The modules contain in addition to the BCM1F components also sensors from
the current monitor BCM1L.
Both systems are shielded with a double-cage structure. The inner cage is 
connected to the ground of the back-end 
readout. The outer cage is connected to the carbon-fibre
support\footnote{The carbon fiber support is not connected to the CMS detector ground.}. 
The two cages are insulated from each other to mitigate 
frequency dependent pick-up effects 
on carbon fibre structures observed elsewhere~\cite{johnson}.
The eight BCM1F modules with their corresponding 
infrastructure were successfully installed 
and tested at the beginning of August 2008.

\section{ First  Measurements with  LHC  Beams}
\label{measurements}
When first beam circulated in the LHC at the beginning of  
September 2008, the
BCM1F was operational and signals from beam-halo particles were recorded.
One of the first beam-generated signals from BCM1F, 
observed on an oscilloscope, is shown
in
Figure~\ref{fig:pulse_scope}.

The readout of the BCM1F modules was triggered by a bunch
pickup detector, BPTX~\cite{bptx}, indicating that a proton bunch crossed
the CMS detector.
A time window of about 500$\ns$ with 
respect to the trigger was recorded by the ADC.  
Figure~\ref{fig:pulse_height}
shows, as an example, the spectrum of signals  
from one of the detector modules. 
The signal size is obtained by integration of the signal pulse
over time.
The distributions for the other channels are very similar.
The signal-to-noise ratios obtained from these distributions
are also shown in Figure~\ref{fig:pulse_height}
for all channels\footnote{Channel one had a faulty cable at the ADC input at the 
time of this measurement.
}.
The values of the signal-to-noise ratio vary between 15 and 25 and are slightly better than
the ones obtained in laboratory measurements before installation.
Since the LHC was filled  with beam in one direction only, 
the beam halo particles should follow this direction.
To demonstrate the capabilities of BCM1F
the signal arrivals times in the flash ADC are measured 
with respect to the BPTX trigger signal.
The distribution of the difference between arrival times of signals from sensors
in different z-planes
but equal azimuthal positions is shown  
in
Figure~\ref{fig:arrivel_time}.

From a Gaussian fit,  a value of 12.4$\ns$ is obtained for the time difference, 
corresponding precisely to the expected time-of-flight of a relativistic
particle between the two BCM1F planes on the $+$z and $-$z side.
The variance of the Gaussian amounts to
1.8$\ns$, leading to an estimate of the single hit timing resolution
of 1.3 ns.

\section{ Conclusions}
\label{conclusions}

The BCM1F is a fully functional sub-detector of the 
BRM system of CMS and will be vital for monitoring beam conditions
close to the beam-pipe inside CMS. It comprises 8 modules each containing 
an sCVD diamond sensor, a front-end ASIC and an optical analog signal transmission
to scalers, flash ADCs and TDCs.
The system
is operated independently 
from the other CMS sub-detectors.                              

System tests of each module in the laboratory show that performance matches requirements.

Samples of sCVD sensors were exposed to a high intensity proton beam 
up to fluences expected at the location of BCM1F for the nominal LHC running. 
The signal amplitude measured for MIPs is reduced to 20\% of the original one,
approaching a level critically low for MIP counting.
Using the ADC the size of the MIP amplitude will be monitored as a function 
of the LHC operation time allowing us to replace the
sensors if necessary.

BCM1F was successfully installed and was operational when LHC 
was filled with first beam. Data taken with beam
show a slightely better signal-to-noise as reached in the laboratory tests. 
A measurement of the signal arrival times using a flash ADC 
indicates 
a
promising
single hit timing resolution of about 
1.3$\ns$. This will allow the separation of incoming halo particles correlated to a certain bunch 
from 
interaction products created at the IP. 

Different readout-modes and time windows for data capture 
are programmable. Data can be written on a local disk 
and published to the CMS readout system. Local data preprocessing
will deliver the shift-crew a detailed picture on the beam-halo count rates 
as a function of the time. Data of several orbits stored in circular memory
will allow diagnostics just after a beam abort.                            
                                                
The BCM1F detector is ready for data taking in the commissioning 
phase of the LHC.

\section{Acknowledgments}

We thank J.P. Chatelain for his contributions 
to design and manufacture the mechanical structures.
We would like to thank our colleagues in the BRM group 
for their advice and assistance. We are grateful to CMS
Technical Coordination for assistance in 
launching and sustaining the project, particularly in its early phases.
We also acknowledge the contribution of the CERN, CMS and DESY 
technical teams which helped bring it to a
successful conclusion. 
We express our gratitude to PSI for allowing us
to use the proton beam and in particular
to K. Deiters for his collaboration there. 
The Rutgers University group is grateful for the support by NSF. R.
Hall-Wilton is grateful for the support 
of the Israeli Technical Associates Program.



\end{document}